\documentclass[superscriptaddress, amsmath,amssymb,notitlepage,aps,twocolumn,prl]{revtex4-2}
\usepackage[latin9]{inputenc}
\usepackage{amsmath}
\usepackage{graphicx}

\makeatletter
\usepackage[colorlinks,linkcolor=blue, urlcolor=blue, anchorcolor=blue, citecolor=blue]{hyperref}
\usepackage{dcolumn}
\usepackage{bm}

\makeatother

\begin{document}
\title{Quantum-enhanced weak absorption estimation with correlated photons}
\author{Zhucheng Zhang}
\affiliation{Graduate School of China Academy of Engineering Physics, Beijing 100193,
China}
\author{Xue Zhang}
\affiliation{Graduate School of China Academy of Engineering Physics, Beijing 100193,
China}
\author{Jing Liu}
\altaffiliation[Present address:~] {Center for Theoretical Physics and School
of Physics and Optoelectronic Engineering, Hainan University,
Haikou 570228, China.}  
\affiliation{National Precise Gravity Measurement Facility, MOE Key Laboratory
of Fundamental Physical Quantities Measurement, School of Physics,
Huazhong University of Science and Technology, Wuhan 430074, China}

\author{Hui Dong}
\email{hdong@gscaep.ac.cn}
\affiliation{Graduate School of China Academy of Engineering Physics, Beijing 100193,
China}
\date{\today}
\begin{abstract}
Conventional absorption spectroscopy relies on coherent laser sources,
and in turn suffers from the inherent limitation of shot noise, especially
in estimating weak absorption. Here we propose a measurement strategy
with correlated photons to determine the weak absorption by distinguishing
the output with and without photons, dubbed as the on-off measurement.
We demonstrate that absorption spectroscopy that incorporates quantum
correlations is capable of estimating weak absorption down to a single-photon
level, even in noisy environments, achieving a precision comparable
to that obtained through 1000 photons in conventional absorption spectroscopy.
Our strategy provides a new method to probe fragile systems with weak
absorption, avoiding the occurrence of light-induced damage.
\end{abstract}
\maketitle
\textit{Introduction}.---Laser spectroscopy, rooted in the subtle
interaction between light and matter, is commonly employed for the
purpose of identifying complex mixtures of chemical compounds and
unraveling the intricacies of chemical reactions with an unprecedented
temporal resolution, diving into the femtosecond realm \citep{RevModPhys.54.697,demtroder1982laser,solarz2017laser,stenholm2012foundations,hannaford2004femtosecond}.
The effectiveness of spectroscopic detection hinges considerably on
the distinctive classical attributes of lasers \citep{yariv1963laser,silfvast2004laser,hecht2010short,shimoda2013introduction},
such as their intensity, coherence, and duration. Conventional spectroscopy
faces inherent limitations due to the shot noise, determined by the
statistical properties of photons \citep{taylor2016quantum}. Increasing
light intensity has been a viable strategy to mitigate shot-noise
effects. However, this strategy inevitably causes light-induced damage
to fragile samples \citep{hopt2001highly,fu2006characterization}.
Utilizing the non-classical properties of lasers \citep{PhysRevLett.98.160401,PhysRevA.79.040305,casacio2021quantum,PhysRevLett.130.133602,PhysRevLett.125.180502},
such as quantum photon correlations, seems to be the only way to break
through these limitations.

It has been well-established that quantum correlations allow more
information to be extracted per photon in optical measurements \citep{slusher1990quantum}.
Such quantum correlations are used routinely to improve the sensitivity
and resolution of phase measurement in the realm of physics, such
as Mach-Zehnder interferometers \citep{PhysRevLett.104.103602,yin2023experimental,PhysRevA.105.032609},
and laser interferometric gravitational wave detectors \citep{aasi2013enhanced}.
Currently, ongoing efforts are devoted to exploring the potential
of quantum correlations in breaking the classical constraints in laser
spectroscopy for investigating the absorption and emission of natural
photosynthetic complexes \citep{li2023single}, increasing the spatial
resolution of microscopy \citep{casacio2021quantum,ono2013entanglement,moreau2019imaging,samantaray2017realization,PhysRevA.98.012121,sabines2019twin},
and improving the measurement precision of material absorption \citep{jakeman1986use,ribeiro1997sub,hayat1999reduction,brida2010experimental,kalashnikov2016infrared,PhysRevLett.125.180502,scarcelli2003remote,PhysRevA.69.013806,moreau2017demonstrating,whittaker2017absorption,losero2018unbiased,okamoto2020loss,PhysRevResearch.6.013034,jonsson2022gaussian,PhysRevApplied.15.044030}.
By exploiting quantum correlations, it is promising to design spectroscopic
techniques that can overcome the classical limitations.

In this letter, we design a new measurement strategy for absorption
estimation that utilizes quantum correlations between photons. Absorption
spectroscopy is a well-used technique to characterize chemical and
biological samples by measuring their absorption of electromagnetic
radiation \citep{butler1964absorption,platt2008differential}. Each
substance has a distinct absorption spectrum that can be utilized
for fingerprint recognition of materials. However, conventional methods
for absorption estimation, which predominantly rely on coherent laser
sources, encounter inherent limitations in precision, especially when
determining weak absorption in noisy environments \citep{PhysRevLett.98.160401,PhysRevA.79.040305}.
Our measurement strategy shares similar configuration to Gaussian
quantum illumination \citep{lloyd2008enhanced,PhysRevLett.101.253601,PhysRevA.80.052310,PhysRevLett.110.153603,PhysRevLett.114.110506},
and such configuration has been employed to achieve loss-tolerant
quantum absorption measurements in a vacuum environment \citep{okamoto2020loss}.
We demonstrate that absorption spectroscopy with quantum correlations
allows us to break inherent limitations via photon counting detection
and the Bayesian estimation, effectively estimating weak absorption
down to a single-photon level within a noisy environment.

\begin{figure}
\includegraphics[scale=0.65]{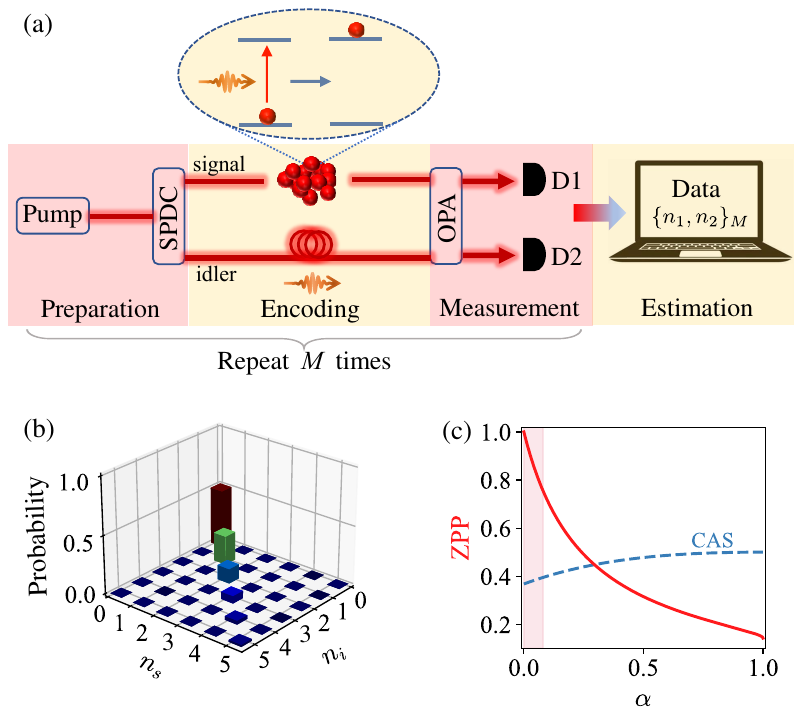} \caption{\protect\label{fig1} Experimental model for quantum absorption spectroscopy.
(a) Correlated signal-idler lights, prepared by spontaneous parametric
down conversion (SPDC) in a nonlinear media, are used to estimate
the absorption coefficient of the sample, in which the signal light
interacts with the sample and the idler light acts as an ancilla.
At the measurement end, the transmitted light and the idler light
undergo an optical parametric amplification process through the utilization
of an optical parametric amplifier (OPA). Additionally, two detectors,
D1 and D2, are employed for joint photon counting. The above measurement
process is then repeated independently $M$ times, yielding a series
of data $\{n_{1},n_{2}\}_{M}$. Finally, these repeated measurement
data are statistically analyzed to estimate the absorption of the
sample. (b) Probability distribution $|C_{n_{s},n_{i}}|^{2}$ of photon
number of the incident signal-idler lights is plotted as functions
of $n_{s}$ and $n_{i}$. (c) Zero-photon probability (ZPP) $P(\{0,0\}|\alpha)$
detected at the measurement end is plotted as a function of the absorption
coefficient $\alpha$, where the counterpart in the conventional absorption
spectroscopy (CAS) is illustrated for comparison. The parameters used
in the numerical simulation include the average number of photons
in the incident signal light, $\bar{n}_{s}=\sinh^{2}|\xi|=1$, and
the environmental mode, $n_{\mathrm{th}}=1$.}
\end{figure}

\textit{Quantum absorption spectroscopy}.---In the spectroscopic
absorption measurements, the sample is typically in an environment
with thermal noise. When illuminating the sample, the incident signal
light, denoted by the creation operator $a_{s}^{\dagger}$, will lose
photons due to the excitation of molecules inside the sample. This
process corresponds to the signal light passing through a thermal
loss channel, which is equivalent to a beam splitter mixing the incident
light with the environmental noise \citep{holevo2007one,RevModPhys.84.621},
namely, 
\begin{equation}
a_{s}^{\dagger}\rightarrow\sqrt{1-\alpha}a_{s}^{\dagger}+\sqrt{\alpha}e^{\dagger},
\end{equation}
where $\alpha\in[0,1]$ is the absorption coefficient of the sample
and $e$ is the environmental mode with average photon number $n_{\mathrm{th}}$
to characterize the intensity of the thermal noise. Then the information
about the sample, i.e., absorption coefficient $\alpha$, is encoded
into the transmitted light. As shown in Fig.~\ref{fig1}(a), our
quantum absorption spectroscopy (QAS)\textbf{ }employs a quantum source
of correlated signal-idler lights, prepared by spontaneous parametric
down conversion (SPDC) in a nonlinear media. The signal light illuminates
the sample, while the idler light serves as an ancilla and does not
directly participate in the interaction with the sample. The two beams
of light are in a two-mode squeezed vacuum state $\ensuremath{|\psi\rangle_{\mathrm{tmsv}}\equiv S(\xi)|0,0\rangle}=\sum_{n_{s},n_{i}}C_{n_{s},n_{i}}|n_{s},n_{i}\rangle$
with
\begin{equation}
C_{n_{s},n_{i}}=\begin{cases}
\dfrac{1}{\cosh|\xi|}\left(-e^{i\theta}\tanh|\xi|\right)^{n_{s}}, & \text{ if }n_{i}=n_{s},\\
0, & \text{ if }n_{i}\neq n_{s},
\end{cases}
\end{equation}
which possess identical photon number distribution \citep{agarwal2012quantum,SM},
as illustrated in Fig.~\ref{fig1}(b), in which $\xi=|\xi|e^{i\theta}$
is the squeezed parameter of the two-mode squeezed operator $S(\xi)=\exp(\xi^{*}a_{s}a_{i}-\xi a_{s}^{\dagger}a_{i}^{\dagger})$
with $a_{i}(a_{i}^{\dagger})$ as the annihilation (creation) operator
of the idler light. By determining the photon number of the idler
light, quantum correlations between them allow one to indirectly infer
the photon number of the signal light. At the measurement end, an
optical parametric amplifier (OPA) is applied to perform a two-mode
squeezing operation via the operator $S(\zeta)$ on the transmitted
and the idler light. Two detectors (D1 and D2) are employed for joint
photon counting in the output signal and idler lights, yielding outcomes
$n_{1}$ and $n_{2}$ with the probability $P(\{n_{1},n_{2}\}|\alpha)$,
conditioned on the absorption coefficient $\alpha$. The probability
$P(\{n_{1},n_{2}\}|\alpha)$ is given by \citep{SM}
\begin{eqnarray}
P(\{n_{1},n_{2}\}|\alpha) & = & \frac{1}{(2\pi)^{2}}\int\mathrm{d}\mathbf{x}\exp\left[-\frac{1}{4}\mathbf{x}^{\mathrm{T}}\left(\mathbf{V}-\mathbb{I}_{4}\right)\mathbf{x}\right]\nonumber \\
 &  & \times F_{1}\left(n_{1}+1;1;-\frac{x_{1}^{2}+x_{2}^{2}}{2}\right)\nonumber \\
 &  & \times F_{1}\left(n_{2}+1;1;-\frac{x_{3}^{2}+x_{4}^{2}}{2}\right),
\end{eqnarray}
where $\mathbf{x}=(x_{1},x_{2},x_{3},x_{4})^{\mathrm{T}}$ is a real
vector, $\mathbf{V}$ is the covariance matrix of quantum state $\rho$
in the output lights with elements $V_{jk}=2\mathrm{Tr}[\rho(R_{j}-d_{j})(R_{k}-d_{k})]-iJ_{jk}$,
$\mathbb{I}_{4}$ is a $4\times4$ identity matrix, and $F_{1}(\bullet;\bullet;\bullet)$
is the hypergeometric function. Here, $R_{j}$ is an element of the
orthogonal operator vector $\mathbf{R}=(q_{s},p_{s},q_{i},p_{i})^{\mathrm{T}}$
with $q_{s,i}=(a_{s,i}+a_{s,i}^{\dag})/\sqrt{2}$ and $p_{s,i}=-i(a_{s,i}-a_{s,i}^{\dag})/\sqrt{2}$.
$d_{j}=\mathrm{Tr}(\rho R_{j})$ is an element of the displacement
vector $\mathbf{d}$, and $J_{jk}=-i[R_{j},R_{k}]$ with $[\bullet,\bullet]$
as the commutator. 

The measurement process is independently repeated $M$ times to obtain
a series of data $\{n_{1},n_{2}\}_{M}$. With these measurement data,
we have to utilize the statistical inference method, Bayesian estimation,
to extract the absorption coefficient \citep{SM}. Using the $i$-th
measurement date $\{n_{1},n_{2}\}_{M}^{(i)}$ from the repeated measurements,
we have the posterior probability distribution $P\left(\alpha|\{n_{1},n_{2}\}_{M}^{(i)}\right)$
through the Bayes\textquoteright{} rule as
\begin{equation}
P\left(\alpha|\{n_{1},n_{2}\}_{M}^{(i)}\right)=\frac{P\left(\{n_{1},n_{2}\}_{M}^{(i)}|\alpha\right)P_{i}(\alpha)}{\int P\left(\{n_{1},n_{2}\}_{M}^{(i)}|\alpha\right)P_{i}(\alpha)\mathrm{d}\alpha},
\end{equation}
where $P_{i}(\alpha)\equiv P\left(\alpha|\{n_{1},n_{2}\}_{M}^{(i-1)}\right)$
is the prior probability distribution obtained from the $(i-1)$-th
measurement. And the prior probability distribution $P_{1}(\alpha)$
in the first measurement is assumed to be uniform in the regime $[0,1]$,
i.e., $P_{1}(\alpha)=1$. After iterating $M$ times, we have the
probability $P(\alpha|\{n_{1},n_{2}\}_{M})$ conditional on the $M$
repeated measurement data $\{n_{1},n_{2}\}_{M}$. The estimated value
$\hat{\alpha}$ of the absorption coefficient is obtained as $\hat{\alpha}=\int\alpha P(\alpha|\{n_{1},n_{2}\}_{M})\mathrm{d}\alpha$,
and the corresponding estimated variance $\delta^{2}\hat{\alpha}$
is determined via the following equation,
\begin{eqnarray}
\delta^{2}\hat{\alpha} & = & \int\alpha^{2}P(\alpha|\{n_{1},n_{2}\}_{M})\mathrm{d}\alpha\nonumber \\
 &  & -\left(\int\alpha P(\alpha|\{n_{1},n_{2}\}_{M})\mathrm{d}\alpha\right)^{2}.
\end{eqnarray}

\begin{figure}[t]
\noindent\includegraphics[scale=0.59]{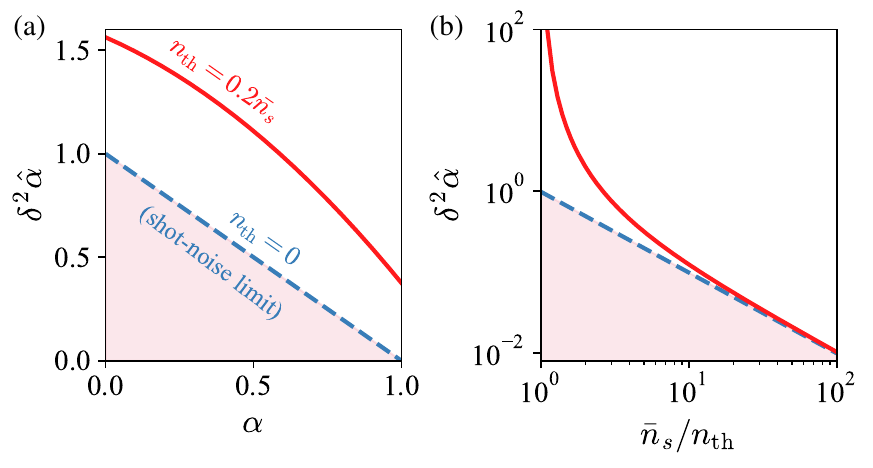}

\caption{\protect\label{Fig_2}Estimated variance $\delta^{2}\hat{\alpha}$
in conventional absorption spectroscopy. Estimated variance as functions
of (a) the absorption coefficient $\alpha$ and (b) the average number
$\bar{n}_{s}$ of input photons. The shot-noise limit is the measurement
precision limit in CAS, i.e., $\delta^{2}\hat{\alpha}=(1-\alpha)/\bar{n}_{s}$,
which can only be achieved when all noise is eliminated. The related
parameters used in the numerical simulation are: (a) $\bar{n}_{s}=1$,
$n_{\mathrm{th}}=0,0.2\bar{n}_{s}$, and (b) $n_{\mathrm{th}}=1$,
$\alpha=0.01$.}
\end{figure}

The conventional absorption spectroscopy (CAS) typically employs an
input of a laser under the coherent state \citep{butler1964absorption,platt2008differential}.
And the average number $\langle a_{s}^{\dagger}a_{s}\rangle_{\mathrm{out}}$
of output photons is measured without photon number counting, namely,
intensity measurement. The estimation is performed by comparing a
known average number $\langle a_{s}^{\dagger}a_{s}\rangle_{\mathrm{in}}$
of input photons and the measured average number of output photons,
i.e., $\hat{\alpha}=1-\langle a_{s}^{\dagger}a_{s}\rangle_{\mathrm{out}}/\langle a_{s}^{\dagger}a_{s}\rangle_{\mathrm{in}}$,
with the estimated variance as $\delta^{2}\hat{\alpha}=\Delta^{2}(a_{s}^{\dagger}a_{s})/|\partial\langle a_{s}^{\dagger}a_{s}\rangle_{\mathrm{out}}/\partial\alpha|^{2}$
\citep{SM}, in which $\Delta^{2}(a_{s}^{\dagger}a_{s})=\langle(a_{s}^{\dagger}a_{s})^{2}\rangle_{\mathrm{out}}-\langle a_{s}^{\dagger}a_{s}\rangle_{\mathrm{out}}^{2}$
is the variance of output photons. Figures \ref{Fig_2}(a) and \ref{Fig_2}(b)
show the estimated variances for CAS as functions of the absorption
coefficient $\alpha$ and the average number $\bar{n}_{s}$ of input
photons, respectively. The dashed lines show the so-called shot-noise
limit in CAS. Given a constant input light intensity, one can see
that the CAS exhibits high estimation variance in the region of weak
absorption, especially in noisy environments, as shown in Fig.~\ref{Fig_2}(a).
Meanwhile, in a noisy environment, the estimated variance tends towards
the shot-noise limit as the number of input photons increases significantly
\citep{PhysRevLett.98.160401}, as illustrated in Fig.~\ref{Fig_2}(b).
Clearly, CAS exhibits limitations in measuring weak absorption samples
in noisy environments.

\textit{Quantum-enhanced sensitivity for weak absorption}.---In contrast
to CAS, the utilization of quantum-correlated light sources offers
inherent benefits for estimating weak absorption samples. The signal-idler
lights generated through the SPDC process in our QAS have a non-classical
photon number distribution {[}see Fig.~\ref{fig1}(b){]}, exhibiting
quantum correlations. Quantum-correlated light sources have been widely
used in optical measurement \citep{PhysRevLett.104.103602,yin2023experimental,PhysRevA.105.032609,aasi2013enhanced}
and imaging \citep{casacio2021quantum,ono2013entanglement,moreau2019imaging,samantaray2017realization,PhysRevA.98.012121,sabines2019twin},
demonstrating significant improvement in sensitivity and resolution.
Used here they allow the enhancement of sensitivity to weak absorption
samples, even in environments where the input signal is disturbed
by the thermal noise. 

To illustrate physical origin of quantum enhancement, we investigate,
in Fig.~\ref{fig1}(c), the change of the probability $P(\{0,0\}|\alpha)$
of detecting zero photons at the measurement end,
\begin{equation}
P(\{0,0\}|\alpha)=\frac{4}{\sqrt{\det(\mathbf{V}+\mathbb{I}_{4})}},
\end{equation}
with the absorption coefficient $\alpha$. During the measurement
process, we utilize OPA to perform a two-mode squeezing operation,
which anti-squeezes the output signal-idler lights under the parameter
regime with $\zeta=-\xi$. In the extreme case of full transmission,
such anti-squeeze results in zero-photon probability (ZPP) for both
detectors, namely, $P(\{0,0\}|\alpha=0)=1$. And the ZPP remains dominant,
ensured by the quantum correlations, in the weak absorption region
of the sample. Such properties allow a simple measurement scheme discussed
later. The sensitivity of the QAS is mainly determined by the change
rate of $P(\{0,0\}|\alpha)$ upon a small change of the absorption
coefficient $\alpha$. And $|\mathrm{d}P(\{0,0\}|\alpha)/\mathrm{d}\alpha|$
is large in the weak absorption region (solid line). On the contrary,
the ZPP in CAS is not sensitive to the change of the absorption coefficient
$\alpha$ (dashed line). Such a picture illustrates that quantum correlations
will enable more information to be gained in the weak absorption region
of the sample, thereby improving the sensitivity of absorption estimation.
The exact explanation for quantum enhancement is presented in section
4 of supplemental material \citep{SM}.

\begin{figure*}[t]
\noindent\includegraphics[scale=0.75]{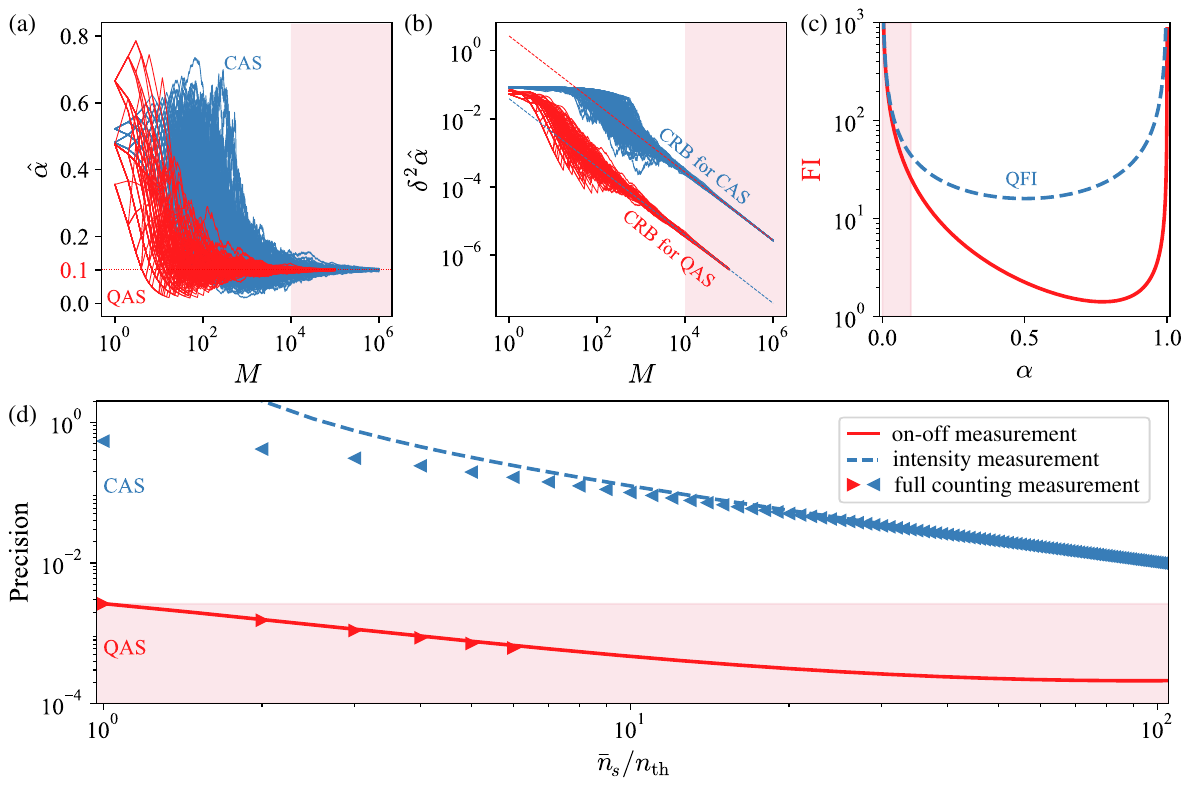} \caption{\protect\label{fig3}Quantum-enhanced weak absorption estimation.
(a)-(b) Estimated value $\hat{\alpha}$ and estimated variance $\delta^{2}\hat{\alpha}$,
associated with the on-off measurement, are plotted as a function
of the number $M$ of repeated measurements in 100 rounds of simulation
experiments, where the measurement data $\{n_{1},n_{2}\}_{M}$ in
each round of experiments are generated by the Monte Carlo simulation
and the true value of the absorption coefficient is set as $0.1$.
(c) Fisher information (FI), associated with the on-off measurement,
is plotted as a function of the absorption coefficient $\alpha$,
where the quantum Fisher information (QFI) is plotted as a comparison.
(d) Estimation precision is plotted as a function of the average number
of photons $\bar{n}_{s}$ in the incident signal light, where the
on-off measurement and the full counting measurement are considered.
To highlight quantum-enhanced estimation precision, the intensity
measurement and full counting measurement in CAS are plotted. The
related parameters used here are: (a)-(c) $\bar{n}_{s}=1$ and $n_{\mathrm{th}}=1$;
(d) $n_{\mathrm{th}}=1$ and $\alpha=0.01$. Here, the estimation
precision based on the full counting measurement is finished with
an open-source toolkit for quantum parameter estimation \citep{PhysRevResearch.4.043057}.}
\end{figure*}

\textit{Quantum-enhanced absorption estimation}.---To demonstrate
quantum-enhanced absorption estimation, we investigate the variation
of estimated value $\hat{\alpha}$ under multiple repeated measurements.
Noticing the fact that the ZPP $P(\{0,0\}|\alpha)$ dominates in the
weak absorption region of the sample, we simplify the measurement
scheme for the two detectors. The two detectors only need to distinguish
whether there is photon counting, dubbed as the \textit{on-off measurement},
yielding four types of measurement outcomes, namely, $\{0,0\}$, $\{0,\emptyset\}$,
$\{\emptyset,0\}$, and $\{\emptyset,\emptyset\}$, with theoretical
probabilities $P(\{0,0\}|\alpha)$,
\begin{eqnarray}
P(\{0,\emptyset\}|\alpha) & = & \sum_{n_{2}=1}^{\infty}P(\{0,n_{2}\}|\alpha),\\
P(\{\emptyset,0\}|\alpha) & = & \sum_{n_{1}=1}^{\infty}P(\{n_{1},0\}|\alpha),\\
P(\{\emptyset,\emptyset\}|\alpha) & = & 1-P(\{0,0\}|\alpha)-P(\{0,\emptyset\}|\alpha)\nonumber \\
 &  & -P(\{\emptyset,0\}|\alpha),
\end{eqnarray}
respectively, in which the symbol $\emptyset$ represents that the
detector has photon counts \citep{SM}. This type of on-off detector
has been successfully utilized in optical phase estimation \citep{PhysRevA.92.053835,PhysRevResearch.1.032024,PhysRevA.105.012607},
significantly reducing the requirement for high-performance detectors
in practical experiments. As shown in Fig.~\ref{fig3}(a), we demonstrate
the variation of the estimated value $\hat{\alpha}$ with respect
to the number $M$ of repeated measurements in 100 rounds of simulation
experiments, in which the measurement data $\{n_{1},n_{2}\}_{M}$
in each round of experiments are generated through Monte Carlo simulation
\citep{SM}. The curves in the figure illustrate that with the input
of correlated signal-idler lights, the estimated value converges asymptotically
to the true value, here $0.1$, in each round of simulation experiments.
Compared with CAS, the introduction of quantum correlations significantly
reduces the number for repeated measurements in each round of experiments
when converging to the true value.

Figure~\ref{fig3}(b) demonstrates the variation of the estimated
variance $\delta^{2}\hat{\alpha}$ with the number $M$ of repeated
measurements. The results show that under the same repeat times as
CAS, quantum correlations guarantee a superior estimation accuracy
in determining the absorption of the sample. Additionally, the curves
further reveal that the estimated variance gradually diminishes in
each round of experiments, ultimately converging towards a specific
bound, as the number of measurements increases. This bound is the
well-known Cram\'er-Rao bound (CRB) in parameter estimation theory \citep{cramer1999mathematical,fisher1923xxi,helstrom1969quantum,holevo2011probabilistic,PhysRevLett.72.3439,paris2009quantum,wiseman2009quantum,liu2020quantum,liu2022optimal},
namely, $1/(MF)$, in which $F$ is the Fisher information (FI) associated
with our on-off measurement scheme (see supplemental material section
4 \citep{SM}), characterizing the extracted information about the
sample absorption. 

\textit{Reaching the ultimate precision limit}.---The performance
of the current on-off measurement scheme is evaluated in Fig.~\ref{fig3}(c)
with FI as a function of $\alpha$, see the solid curve. The large
value of FI at the small absorption side demonstrates that quantum
correlations enhance sensitivity for the weak absorption samples in
noisy environments. An ultimate upper bound for FI exists and is termed
quantum Fisher information (QFI) $\mathcal{F}$, with the explicit
form for our QAS as (see supplemental material section 5 \citep{SM})
\begin{equation}
\mathcal{F}=\frac{\bar{n}_{s}+n_{\mathrm{th}}+2\bar{n}_{s}n_{\mathrm{th}}}{\alpha(1-\alpha)}.
\end{equation}
In the weak absorption region, the FI in our QAS is only different
from the ultimate upper bound with a small deviation. It is clear
that our on-off measurement scheme achieves the ultimate precision
limit in estimating the weak absorption in noisy environments, similar
to the findings in Ref. \citep{okamoto2020loss}, where their sensitivity
attains the quantum limit in a vacuum environment.

The on-off measurement scheme is capable of estimating the weak absorption
coefficient with a limited number of photons. In Fig.~\ref{fig3}(d),
the estimation precision, $\delta^{2}\hat{\alpha}$, is plotted as
a function of the average number of photons $\bar{n}_{s}$ of the
incident signal light for both QAS (solid curve) and CAS (dashed curve).
The results show clear advantages of QAS over CAS in estimating the
weak absorption. We demonstrate that our on-off measurement scheme
allows the estimation of the weak absorption down to a single-photon
level within the noisy environment. And an average one-photon correlation
source can achieve the estimation precision comparable to that achieved
with about 1000 photons in CAS. The low photon number exploited in
the current on-off measurement scheme allows us to avoid the light-induced
damage to the sample, compared with the common strategy of increasing
light intensity to mitigate the shot noise in CAS \citep{butler1964absorption,platt2008differential},
while achieving comparable estimation precision. 

One potential improvement for precision is to utilize the full counting
measurement \citep{couteau2023applications,cahall2017multi,PhysRevA.63.033812,hadfield2016superconducting,banaszek2003photon},
where the exact photon number is detected to reveal the full distribution
$P(\{n_{1},n_{2}\}|\alpha)$, allowing for maximum information extraction
\citep{SM}. In Fig.~\ref{fig3}(d), we further show the estimation
precision obtained with the full counting measurement. We demonstrate
that our QAS (right triangle) offers more information on the sample,
enabling a superior level of estimation precision over CAS (left triangle).
Importantly, the precision in our QAS achieved through the on-off
measurement is also better than that in CAS with the full counting
measurement.

\textit{Summary and discussion}.---By reporting absorption spectroscopy
with quantum-enhanced sensitivity and using it to improve estimation
precision for samples with weak absorption in noisy environments,
our study quantitatively demonstrates the long-recognized potential
of quantum correlations to overcome classical limitations in spectroscopy.
Our implementation within the on-off measurement scheme provides the
capacity to reach the ultimate precision limit in noisy environments
in weak absorption regions. We show that absorption spectroscopy that
incorporates quantum correlations can measure the weak absorption
down to the single-photon level in noisy environments, with precision
comparable to that achieved by 1000 photons in CAS. This quantum absorption
spectroscopy represents a pivotal application to overcome the classical
constraints on sensitivity and precision of existing high-performance
laser spectroscopy.

Building upon considerations of thermal noise in the sample environment,
we analyze the impact of the OPA and its anti-squeezing operation
on the estimated variance of the weak absorption coefficient, see
section 6 of our supplemental material \citep{SM}. The results show
that the best estimated variance is obtained around the point where
the applied anti-squeezing balances the initial SPDC squeezing. Such
choice of squeezing parameter ensures that the ZPP is sensitive to
the weak absorption coefficient $\alpha$, allowing our on-off measurement
to extract more information about the weak absorption. In addition,
we also analyze the impact of factors such as the loss of the input
light source and the dark count of the detector on the accuracy improvement
brought by quantum correlations. We demonstrate that our on-off measurement
scheme remains effective in QAS despite the influence of these factors,
enabling us to break through the shot-noise limit of CAS.
\begin{acknowledgments}
We thank Dr.~Lei Shao for helpful discussions. This work is supported
by the Innovation Program for Quantum Science and Technology (Grant
No.~2023ZD0300700), and the National Natural Science Foundation of
China (Grant Nos.~U2230203, U2330401, 12088101, 12347123, 12405012,
and 12175075).
\end{acknowledgments}

\bibliographystyle{apsrev4-2}
\bibliography{QAbsortption}

\end{document}